# Automatic In-line Quantitative Myocardial Perfusion Mapping: processing algorithm and implementation

**Hui Xue[1], Louise A. E. Brown[2], Sonia Nielles-Vallespin[3], Sven Plein[2], Peter Kellman[1]**

1. National Heart, Lung and Blood Institute, National Institutes of Health, Bethesda, MD, USA
2. Multidisciplinary Cardiovascular Research Centre (MCRC) & Leeds Institute of Cardiovascular and Metabolic Medicine, University of Leeds
3. Royal Brompton Hospital, London, UK

## Corresponding author:

Hui Xue

National Heart, Lung and Blood Institute
National Institutes of Health
10 Center Drive, Bethesda
MD 20892
USA

Phone:  +1 (301) 827-0156
Cell:   +1 (609) 712-3398
Fax:    +1 (301) 496-2389
Email: hui.xue@nih.gov

**Word Count: 5,960**

hui.xue@nih.gov
L.Brown1@leeds.ac.uk
s.nielles@googlemail.com
S.Plein@leeds.ac.uk
kellmanp@nhlbi.nih.gov

## Abstract

**Purpose**

Quantitative myocardial perfusion mapping has advantages over qualitative assessment, including the ability to detect global flow reduction. However, it is not clinically available and remains as a research tool. Building upon the previously described imaging sequence, this paper presents algorithm and implementation of an automated solution for inline perfusion flow mapping with step by step performance characterization.

**Methods**

Proposed workflow consists of motion correction (MOCO), arterial input function (AIF) blood detection, intensity to Gd concentration conversion, and pixel-wise mapping. A distributed kinetics model, blood tissue exchange model (BTEX), is implemented, computing pixel-wise maps of myocardial blood flow (MBF, ml/min/g), permeability-surface-area product ($PS$, ml/min/g), blood volume ($V_b$, ml/g) and interstitial volume ($V_{isf}$, ml/g).

**Results**

N=30 healthy subjects (11 men, 26.4±10.4 yrs) were recruited and underwent adenosine stress perfusion CMR. Mean MOCO quality score was 3.6±0.4 for stress and 3.7±0.4 for rest. Myocardial Dice coefficients after motion correction were significantly improved (P<1e-6), 0.87±0.05 for stress and 0.86±0.06 for rest. AIF peak Gd concentration was 4.4±1.3 mmol/L at stress and 5.2±1.5 mmol/L at rest. Mean MBF at stress and rest were 2.82±0.47ml/min/g and 0.68±0.16ml/min/g. $PS$ was 1.32±0.26ml/min/g and 1.09±0.21ml/min/g (P<1e-3). Blood volume was 12.0±0.8ml/100g at stress and 9.7±1.0ml/100g at rest (P<1e-9), indicating good adenosine vasodilation response. $V_{isf}$ was 20.8±2.5ml/100g at stress and 20.3±2.9ml/100g at rest (P=0.50).

**Conclusions**

An inline perfusion flow mapping workflow is proposed and demonstrated on normal volunteers. Initial evaluation demonstrates the fully automated proposed solution for the respiratory motion correction, AIF LV mask detection and pixel-wise mapping, from free-breathing myocardial perfusion imaging.

# Introduction

Myocardial perfusion can be evaluated with dynamic cardiovascular magnetic resonance (CMR) during the passage of a contrast agent bolus. Most commonly, perfusion images are evaluated qualitatively, while quantitative evaluation would be more desirable. The potential benefits of quantification include: objective assessment, simpler and faster analysis, and the ability to detect disease with a global reduction in flow such as balanced multi-vessel obstruction or microvascular disease. The desired output of a quantitative perfusion study is a map of myocardial blood flow (MBF) in units of ml/min/g.

Quantification of myocardial perfusion using CMR was first proposed over 20 years ago [1,2], yet qualitative interpretation of images remains the primary means available to clinicians. At the same time, there has been considerable technical development in methods for quantifying myocardial perfusion. The kinetics of gadolinium (Gd) contrast agent have been studied [3–5] and a number of models of myocardial tissue have been proposed. In the category of compartmental models, the exponential [6], Fermi function [7], or BSpline based model free deconvolution [8] have been applied to myocardial perfusion. More comprehensive distributed parameter models have been applied to the estimate of myocardial blood flow in MRI [9–13] and in PET [14–16] based on modeling of the underlying physiology. As illustrated in prior publications [6,11,17–24], the assumption behind the simpler compartmental models is that Gd delivery to the myocardial interstitial space from the vascular space is flow limited, at least at the low flow scenario. In this case, estimates of myocardial blood flow are well approximated using deconvolution methods. On the other hand, additional studies [9,12,13] have suggested that Gd delivery to the myocardium may not always be flow limited, especially under the stress condition. Since the compartmental model assumes a spatially invariant distribution or instantaneous mixing of Gd concentration [6,11] and does not explicitly estimate the influence of Gd extracted from the vascular space into the interstitial space [25], more general distributed parameter models may be preferable and have attracted interest in recent years [12,13,26]. Additional to the estimation of myocardial blood flow (MBF) (ml/min/g), distributed models [9,12,13] offer estimation of other parameters including permeability-surface-area product ($PS$, ml/min/g), blood volume $V_b$ (ml/g) or plasma volume $V_p$ (ml/g), and interstitial volume $V_{isf}$ (ml/g). Additional parameters such as extraction fraction $E$ [6] and capillary transit time $T_c$ (sec) [27] may be derived from these model parameters. These additional parameters characterize myocardial microvascular structures and may have potential diagnostic value [28].

Accuracy of perfusion quantification highly depends on the correct measurement of the arterial input function (AIF). Since longer saturation time leads to saturated signal intensities in perfusion imaging during the contrast uptake, either "dual-bolus" [29] or "dual-sequence" [30] techniques have been proposed for more accurate AIF estimation. The former relies on injecting a separate low dose contrast bolus for AIF estimation and assumes signal linearity between contrast concentration and signal intensity at this dose. The latter modifies the saturation recovery sequence to acquire low resolution images (so-called AIF images) with both very short saturation time and echo time, reducing signal saturation. Several important studies have proven the utility of perfusion quantification using the "dual-bolus" method [31]. The "dual-sequence" technique requires only one contrast bolus injection and therefore simplifies the clinical workflow. Using the dual-sequence method, the assumption of signal linearity to Gd concentration can be removed by converting the signal intensity of AIF and perfusion images to [Gd] (Gd concentration unit, mmol/L). This strategy was previously proposed with FLASH perfusion imaging sequences [32,33]. Our recent development of the dual-sequence technique [34] further extended AIF imaging to acquire two echoes during every readout and correcting for the signal loss due to the shortened T2* at high contrast concentrations.

Previously reported implementations of perfusion quantification [8,19,35] have been off-line and time consuming, which limits the application of quantitative perfusion in a clinical setting. As a result, quantitative perfusion CMR has remained a research tool. To overcome this limitation, we developed and evaluated a fully automatic, in-line solution for pixel-wise mapping of myocardial blood flow based on our previously described optimized dual sequence method, which can be integrated into a clinical workflow [34].

A number of technical challenges have to be overcome to achieve reliable automated perfusion flow mapping. A typical perfusion scan often lasts 60 or more heart beats. For this long period, patients are unable to hold their breath. Respiratory motion, therefore, must be corrected to allow pixel-wise flow mapping. Other automated processing steps include segmentation of the left ventricular blood pool to estimate the AIF, surface coil inhomogeneity correction, and perfusion mapping based on pixel-wise fitting to a tissue model. Our solution includes: a) automated respiratory motion correction for both AIF and perfusion image series, b) conversion from image intensities to Gd concentration units, c) arterial input function (AIF) LV blood-pool mask detection to compute the input function and d) pixel-wise perfusion mapping to estimate MBF and other parameters. Because distributed perfusion models better approximate myocardial capillary physiology and do not assume complete Gd extraction, a distributed kinetics perfusion model, called the blood tissue exchange model (BTEX) [36,37] was implemented in this study. The solution was implemented in C++ via the Gadgetron streaming image reconstruction software framework [38] which provides a flexible system for creating streaming data processing pipelines where data pass through a series of modules or "Gadgets" from raw data to reconstructed images. These reconstructed images are reinserted into the scanner image reconstruction pipeline, allowing seamless integration of the proposed solution on the MR scanner. As a result, the image reconstruction, AIF signal estimation, respiratory motion corrected perfusion images and myocardial blood flow maps were calculated without any user interaction at the completion of the perfusion MR scan. Our proposed method has recently been compared against PET and demonstrated very good agreement [39].

The imaging sequence used in study was previously published in [34]. This early publication also presented conversion from image intensity to Gd concentration conversion supported by Gd phantom calibration experiments. With the focus on imaging sequence and experimental setup, the previous paper did not present the full algorithm to compute pixel-wise perfusion maps from [Gd] signals. This paper presents algorithms for each processing step with additional validation results: (1) motion correction using iterative scheme based on KL transform, (2) AIF blood detection using k-means clustering, and (3) pixel-wise perfusion mapping using a coarse-to-fine computational scheme for implementing the partial differential equation (PDE) based BTEX model. Perfusion mapping is performed on N=30 normal healthy volunteers, and histograms of each parameter estimates are examined.

## Methods

### Overview of inline perfusion mapping

The inline perfusion mapping proposed in this paper used the dual-sequence single-bolus injection for simplified clinical workflow [34]. The diagram of Figure 1 is an overview of the main processing steps, with each processing step explained in this section. Low resolution AIF and higher resolution myocardial images were acquired using a 2-D multi-slice saturation recovery sequence. Images were acquired during free-breathing, therefore, motion correction was used to correct in-plane motion. The AIF sequence used two echoes in order to measure and correct for T2* loss during the first pass. To convert signal intensity to units of contrast agent concentration [Gd] for both AIF and myocardial images, proton density (PD) images were acquired before the saturation recovery readouts and used to correct surface coil inhomogeneity and to normalize signal intensity thereby enabling LUT (look-up-table) conversion [34]. Gadolinium concentration signals for the AIF and myocardium were temporally resampled to 0.5 sec per sample using linear interpolation which also corrected for missed triggers resulting in a fixed sampling corresponding to a heart rate of 120 bpm which are then input to the pixel-wise calculation of myocardial blood flow. This also allowed calculation of myocardial blood flow to delay the AIF in 0.5 sec steps in estimating the arterial delay. It was empirically determined that 0.5 sec step size was adequate by viewing the mean square error function vs arterial delay.

*Image acquisition and reconstruction*

The sequence parameters for the dual sequence myocardial perfusion imaging used in this study were as previously described [34] where detailed imaging parameters may be found. Some key imaging parameters are: FOV 360×270 mm$^2$, for myocardial perfusion imaging, matrix size 192×111, interleaved acceleration R=3, Flash readout, TR= 2.1 ms, slice thickness 8 mm, trigger delay TD=72 ms. For AIF, matrix size 64×34, interleaved acceleration R=2, Flash readout, TR=2.45ms, slice thickness 10 mm and TD=2.8 ms. For the perfusion imaging, linear phase encoding order was used with the truncated lines of k-space in latter half. The slice order is from base to apex with the AIF slice for basal slice acquired following the R-wave trigger (TD=2.8 ms). Total imaging duration including SR preparation and delay was 143 ms per slice which allowed imaging the AIF and 3 slices up to high heart rate of 120 bpm. A chemical shift fat saturation was used for the myocardial imaging slices. The total number of measurements including 3 PD weighted frames was typically 60. The bolus was administered at approx. 8 heart beats after the start of the scan to ensure an adequate number of baseline images prior to contrast arrival. The study was performed using the 3T MAGNETOM Prisma, Siemens AG Healthcare, Erlangen, Germany, and used a FLASH protocol [34].

Image reconstruction and processing was implemented using the Gadgetron software framework [38,40,41]. Multichannel data were acquired with temporally interleaved sampling and were noise pre-whitened using pre-scan noise. Parallel imaging reconstruction was performed using TGRAPPA [42]. For the AIF images, parallel imaging kernels are computed on the first echo and applied to both echoes. Raw filtering was used to suppress Gibb's ringing [34,43]. A Gaussian raw filter truncated at 1.5 standard deviations was used. The loss in spatial resolution was 18% compared to uniform weighting, and $1^{st}$ sidelobe was suppressed by >12:1. SNR unit reconstruction [44] was used throughout all processing steps. This ensures optimal SNR in reconstruction images after noise pre-whitening and identical scaling ratio was applied to both saturation recovery and proton density images. Therefore, the normalized SR/PD images can be correctly computed for image intensity to [Gd] conversion. SNR scaling facilitates threshold based noise masking.

*Motion correction (MOCO)*

Free-breathing perfusion images were corrected for respiratory motion using MOCO. This step utilized non-rigid image registration [45,46] applied in an iterative fashion. To cope with significant image contrast variation during the contrast bolus passage, instead of directly registering perfusion images against each other, synthetic perfusion series were derived from a Karhunen–Loève (KL) transform which computed principal eigen-images with similar contrast (KL transform, when applied to discrete random vectors, is computed by principal component analysis, PCA). MOCO was achieved by registering perfusion images pairwise with KL series. As shown in Figure 2, the KL transform was computed over a sliding temporal window with variable width. Suppose the number of perfusion images was N, the initial temporal window width of KL transform was N/3. To gradually bring back perfusion contrast, this process was iteratively repeated. At each iteration, the synthetic images were recomputed from the registered perfusion series with a decreased window width. This approach can deal with significant contrast changes during the bolus passage which had been shown to be a major challenge for myocardial perfusion motion correction [47]. This algorithm iterated non-rigid image registration and KL transform based model image estimation to decouple perfusion contrast changes from respiratory motion. One example of this iterative MOCO is shown in Figure 3. This algorithm was applied to both perfusion and AIF series. By using non-rigid registration, the iterative MOCO was applied to the entire image without need to crop out the heart region. More details are given in the Appendices.

*AIF LV blood detection*

The motion corrected low resolution AIF image series was used to extract the arterial input function. Figure 4 illustrates the processing steps for automatically detecting the AIF LV blood pool. First, the AIF PD image was used to detect the noise background. Since the noise standard deviation (SD) was unity after the SNR unit reconstruction, a simple threshold of 3 SD's was used. For all foreground pixels as determined by the noise mask, the time intensity curves were analyzed and foot and peak time points were determined using a scale-space based detector [48]. The foot time point was defined as the moment of contrast arrival and the time of peak was the moment where contrast concentration reached its maximum value (illustrated in Figure 4). To achieve robust detection in the presence of additive noise, the scale-space detector first creates multiple smoothed AIF curves, filtered with a Gaussian kernel with different values of sigma as "scale". If a feature, such as the AIF peak point, can be detected on the set of smoothed curves across all scales, it was considered a consistent feature. That is, this strategy achieved detection in both temporal and scale dimension. If a feature point existed in the smaller scale and vanished in the larger scale, it was not a stable feature, but rather a local signal change or noise-caused feature. After detecting the foot and peak time points, the upslope, area-under-curve (AUC), and peak time were computed for every foreground pixel. The blood pool has very strong contrast uptake and therefore high upslope and AUC value. Pixels with values in both the top 10% upslope and top 10% AUC were picked as the candidates for LV blood pool mask. A secondary classification of all candidate pixels was based on statistical clustering computed using the k-means algorithm [49] with 2 initial clusters. The 2 initial clusters were classified as RV and LV based on arrival time, and the LV cluster was re-clustered using k-means with 4 clusters. The LV cluster was selected as the one with the highest correlation coefficient between the centroid signal of each candidate cluster with the original LV cluster (Figure 4). The AIF image, due to its lower spatial resolution, is more vulnerable for partial volume effects. Edge pixels can often have reduced intensity values due to partial volume with adjacent tissue and limited spatial resolution. The final LV blood pool mask was calculated using a further erosion step based on keeping the top 15% percentile values.

*Conversion from signal intensity to gadolinium concentration*

The dual-echo AIF signals were extracted from the blood pool mask and used to correct the T2* signal loss, as proposed in [34]. Both the AIF and the myocardial images were corrected for surface coil intensity variation using the initial PD frames which had been co-registered with the saturation recovery images. To convert signal intensity to [Gd], the measured AIF and myocardial perfusion SR images were normalized by the intensity of proton density images to get the SR/PD value as input to the LUT. As described in [34], a Bloch simulation was performed to compute the readout magnetization of SR and PD images. Note that the results varied slightly from previously reported [34] due to improved modeling of the slice profile introduced more recently. The previous paper used a simple uniform slice profile for readout pulse, whereas in this paper the improved software incorporated the actual Hanning weighted sinc RF pulse into the Bloch equations. The inline mapping software extracted all necessary parameters on the fly from the actual imaging protocol, such as saturation recovery delay time (TD), number of phase encodes, acceleration factor, readout sequence type (FLASH or SSFP) and flip angles etc. A look-up table was constructed with the horizontal axis being the [Gd] (0 to 20 mmol/L with step size of 0.01mmol/L), and the vertical axis being the normalized intensity SR/PD (SR signal intensity normalized by PD signal intensity). Separate LUTs were constructed for the low resolution AIF and the higher resolution myocardial imaging protocols.

*Perfusion mapping*

Perfusion Gd images and AIF Gd curve were input into the distributed blood tissue exchange (BTEX) model [36,50] for the estimation of myocardial blood flow and other parameters. For every pixel in the heart region, BTEX model solved two partial differential equations:

$$\frac{\partial C_p}{\partial t} = \frac{-F_p L}{V_p} \cdot \frac{\partial C_p}{\partial x} + \frac{PS}{V_p} \cdot (C_{isf} - C_p) + D_p \cdot \frac{\partial^2 C_p}{\partial x^2} \quad (1)$$

$$\frac{\partial C_{isf}}{\partial t} = -\frac{PS}{V_{isf}} \cdot (C_{isf} - C_p) + D_{isf} \cdot \frac{\partial^2 C_{isf}}{\partial x^2} \quad (2)$$

where subscripts $p$ and $isf$ corresponded to plasma and interstitial fluid space, respectively. $C$ was contrast agent concentration. Four parameters to be estimated were: $F$, blood flow, $PS$, permeability surface area product, $V_p$ and $V_{isf}$, plasma and interstitial volume. $D_p$ and $D_{isf}$ were the Gd molecular diffusion coefficients within the vascular and interstitial space, set to be fixed at 1e-5cm$^2$/sec and 1e-6cm$^2$/sec [37,51]. L was the capillary length, fixed to be 1 mm, as suggested in [36]. The total length L was divided into a finite number of steps (30 steps were used) and PDEs were solved on the grid. The hematocrit HCT was required to convert blood [Gd] concentration to plasma concentration for AIF and used in calculating final blood flow and blood volume:

$$C_p(t) = \frac{C_b(t)}{1-HCT}, \quad F_b = \frac{F_p}{1-HCT}, \quad V_b = \frac{V_p}{1-HCT} \tag{3}$$

A fixed value of HCT = 0.42 was assumed throughout for the tested normal subjects, and sensitivity to this HCT parameter was analyzed later in the results. Myocardial density used in this study was 1.05 g/ml [52].

Pixel-wise BTEX model fitting was solved iteratively, as illustrated in Figure 5. The fitting process starts with an initial guess of model parameters. The BTEX model parameters were then solved for using the AIF [Gd] curve as the driving input function. The resulting Gd residual signal for each set of parameters was compared to the measured perfusion [Gd] curve and the parameter estimate used the mean squared error (L2 norm) as the goodness-of-fit measure. Model parameters were then adjusted in an optimization step until convergence. More details for BTEX modelling are given in Appendices.

***Inline integration***

To deploy the proposed solution in a clinical setting, all processing steps were fully automated. Software was implemented using C++ on the Gadgetron framework [38,40]. Gadgetron software may be run using several configurations including: on the scanner image reconstruction computer, on a networked computer, or using cloud computing [40]. In this study, an external networked PC was used, and all raw data was saved to enable retrospective analysis. Raw data was converted to the ISMRMRD standard [41] which was de-identified and sent to Gadgetron for processing. All processing steps including parallel image reconstruction, motion correction, and pixel-wise flow mapping were performed with OpenMP based multi-threading. Reconstructed perfusion images in intensity and [Gd] units, AIF [Gd] curve plots, and MBF maps were sent back to the scanner host from the Gadgetron computer without any user interaction and all series were saved into the Dicom database. This scanner integration was demonstrated in Figure 6(a) which shows the actual screenshot of the scanner. This figure illustrated a scan using the proposed inline quantitative perfusion mapping. Besides the flow maps and the MOCO images, the AIF signals and measured heart rate during acquisition were also sent back to scanner as plots (Figure 6b). Parametric maps were displayed with custom colormaps. A version that ran directly on the scanner's image reconstruction computer was also tested and reconstruction times were measured for both external PC and using scanner computer hardware.

**In-vivo imaging experiments**

The proposed inline perfusion flow mapping technique was implemented and deployed at Leeds Teaching Hospitals, UK using a 3T clinical MR scanner (Magnetom PRISMA, Siemens, software version VE11C). Identical cardiac perfusion imaging protocols were used for stress and rest. The study was approved by the respective local Institutional Review Board (IRB) and Ethics committee, and all subjects gave written informed consent. Anonymized data was analyzed at NIH with approval by the NIH Office of Human Subjects Research OHSR (Exemption #13156). The Gadgetron based imaging reconstruction was utilized to compute pixel-wise perfusion flow maps using a networked Linux PC based configuration.

N=30 healthy normal volunteers (11 men and 19 women, mean age 26.4±10.4 yrs) were recruited to receive stress and rest perfusion scans. The recruited volunteers had no medical history of diabetes, hypertension, hypercholesterolemia, or bradycardia (<45 heart beat per minute) and had systolic blood pressure >= 90mmHg. All subjects were instructed not to take in caffeine within 24 hours prior to the examination. Gadolinium contrast agent (Gadovist, Bayer Schering Pharma AG) was administered as a bolus of 0.05 mmol/kg at 5 ml/sec with 20 ml saline flush using a power injector (Medrad MRXperion Injection System, Bayer). For stress perfusion, adenosine was administered by continuous intravenous infusion for 4 min at a dose of 140 μg/kg/min before contrast injection. Blood pressure (BP) and heart rate (HR) were recorded during adenosine infusion to monitor haemodynamic response, and symptoms were recorded. Stress images were reviewed for the presence of splenic switch-off to ensure adequate adenosine response [53].

All scans utilized the FLASH perfusion readouts. The stress perfusion was performed first and the rest images were acquired after ~15 mins. All perfusion studies acquired three short axis slices (basal, medial and apical) for every heartbeat. A total of 60 heart beats were imaged. Imaging experiments were conducted completely with free breathing.

**Analysis**

*AIF LV Blood detection*

The AIF LV blood detection step was validated against manual delineation. For manual delineation, an experienced operator (HX, 9 years of experience) drew a ROI in the LV blood pool of the AIF images. Resulting dual-echo time intensity curves went through the same percentile based filtering step and T2* correction as the automated algorithm and was finally converted into [Gd] units. To validate the accuracy of LV blood pool detection, the AIF curves derived from automatic detection was compared to the results using the manually drawn ROI. AIF peak Gd concentration, first-pass duration (from foot to valley, Figure 4) and area-under-cure during first-pass was computed for both auto and manual curves. The automated AIF Gd curves were used to compute flow maps which was compared to maps computed with manual curves. We further visually inspected the masks to evaluate whether the LV was correctly selected in all cases.

*Motion Correction*

Motion correction performance was validated by visual assessment (HX, with 9 years of experiences in perfusion imaging and motion correction) using a score between 1 and 4 (0.5 increment). All motion corrected perfusion series were converted to movie files and viewed in random order. A score of 1 indicated the worst quality and a score of 4 was the best. Specifically, a score of 4 (excellent) was given if motion correction removed all discernible motion and the heart was perfectly still. A score of 3 (good) was given for images with a small amount of residual motion, but still suitable for whole myocardium perfusion quantification. A score of 2 (fair) was given to images with borderline motion correction, which could still be used to quantify blood flow in part of the myocardium. A score of 1 (poor) meant insufficient image quality to perform flow mapping for all myocardium with visible stretching and other motion correction failures.

To quantify the performance of motion correction, myocardium was further manually delineated for these subjects at peak inspiration and expiration. After motion correction, the segmented myocardium was propagated to the corrected images using the deformation fields. An ideal motion correction should lead to perfect overlap between segmented myocardium from two frames. Therefore, the overlap rate before and after motion correction was computed as the Dice similarity coefficient (Dice) [54]. For two segmented regions A and B, the Dice is defined as 2×area(A∩B)/(area(A)+area(B)). This value will be 1 for a perfect overlap and 0 for non-overlap. In addition, the false positive (FP) and false negative (FN) errors were computed. FP was defined as the percentage area of segmented myocardium in the first frame that was not labeled in the second and FN was defined as the percentage area of myocardium in the second that was not labeled in the first. Because the cardiac motion can be non-rigid in its nature, the myocardium boundary errors (MBE) which was defined as the mean distance between endocardial and epicardial contours of two frames were computed for all series as well. The binary mask of myocardium was upsampled by 2x and all boundary pixels were used to compute MBE. While Dice ratio will capture the bulk motion due to failed breath-holding, MBE could highlight the local myocardial deformation. For a perfusion series, these four measures (Dice, FP, FN, and MBE) were computed.

*Perfusion flow mapping*

Pixel-wise maps were analyzed for each slice at both stress and rest with the whole myocardium segmented manually. Endocardial and epicardial borders were drawn, excluding papillary muscles. Where the left ventricular outflow tract (LVOT) was included, or partial volume effect meant myocardium were too thin to contour, these were excluded from further analysis.

To study the influence of actual HCT on perfusion flow values, the HCT used in the BTEX flow mapping was varied from 0.3 to 0.6 in steps of 0.05 and the resulting flow estimates were compared to the flow values for the normal subjects using the nominal hematocrit.

*Statistical Analysis*

The resulting values were presented as mean +/- standard deviation. The paired t-test was used when appropriate, e.g. to compare MBF values from stress and rest studies for the same subject. A P-value less than 0.05 was considered statistically significant. Histograms were calculated for parameter values across 30 subjects.

# Results

For the cohort of N=30 normal subjects, the rest scans were performed 15.5±3.1 mins after stress. The heart rate was 94±13 bpm for stress and 63±8 bpm for rest. The Rate Pressure Product (RPP) was 10776 ± 2296 for stress and 7235 ± 1453 for rest. Perfusion images were visually checked for splenic cutoff to confirm stress response. All MBF maps were first visually inspected to confirm good quality for all subjects. Results of processing steps are reported in following paragraphs.

*Motion correction*

Figure 7 demonstrates typical performance of MOCO (also in Supporting information video S2-3). Given the complete free-breathing acquisition, the automated motion correction "froze" the heart so that the tissue appears stationary (Figure 7a-b), therefore enabling pixel-wise flow mapping. An example of dual-echo MOCO results for AIF images are shown in Supporting information video S4.

The KL based motion correction algorithm was robust in both stress and rest scans to handle contrast changes during the Gd uptake. Mean quality score for motion correction was 3.6±0.4 for the stress and 3.7±0.4 for the rest. To illustrate the performance of the MOCO processing step, Supporting information video S5 and 6 give examples with scores being 4.0 and 3.0. The MOCO example in S6 video was challenging due to significant respiratory motion where the diaphragm experienced large anatomical deformation between frames. The MOCO algorithm is optimized to correct respiratory motion for the heart region and for this case, it was not as effective in aligning the diaphragm where there is drastic through-plane motion. Although the heart was better aligned there is still some residual motion visible in the video.

For the stress, the Dice coefficients were significantly improved after the MOCO processing step (before: 0.67±0.16; after: 0.87±0.05, P<1e-6). The myocardium boundary errors were significantly reduced from 2.28±0.93mm to 0.88±0.19mm (P<1e-6). Same improvements were found for both FP (before: 0.33±0.17; after: 0.12±0.06, P<1e-6) and FN (before: 0.33±0.16; after: 0.13±0.06, P<1e-6). For the rest, before and after the MOCO, Dice was 0.66±0.17 and 0.86±0.06; FP were 0.34±0.18 and 0.13±0.07 and FN were 0.35±0.17 and 0.15±0.07. The myocardium boundary errors were 2.11±0.94mm and reduced to 0.89±0.33mm. All improvements were significant (P<1e-6). Given the acquired spatial resolution of 360mm/192=1.875mm, the residual myocardial boundary errors were under half pixel.

*AIF Gd concentration*

Using the automated AIF blood masking, the AIF peak Gd concentration was 4.4±1.3 mmol/L at stress and 5.2±1.5 mmol/L at rest. The duration of contrast first pass (from foot to valley of AIF time-Gd curve, Figure 4) was 10.2±1.6s at stress and 13.9±2.5s at rest. The T2* in LV blood pool at peak concentration of bolus arrival was 14.7±3.2 ms and 10.6±2.0 ms for stress and rest peak Gd concentration, respectively. Without T2* correction, AIF peak [Gd] decreased to 4.0±1.1 mmol/L (P<1e-3) at stress and 4.7±1.2 mmol/L (P<1e-3) at rest. The pixel-wise MBF maps were computed using both the T2* corrected AIF and the first echo signal curve without T2* correction for comparison. Lack of T2* correction lead to significant overestimation of MBF of 9.1% (P<1e-6), because of the reduction of AIF [Gd] signal.

*AIF blood detection*

Visual inspection verified the auto-generated AIF masks properly detected the LV blood pool for all scans. The manual masking of AIF blood pool gave an AIF peak Gd of 4.2±1.2 mmol/L at stress and 5.2±1.6 mmol/L at rest. First pass duration was 11.0±2.3 and 14.9±3.2 seconds at stress and rest, respectively. AUC was 17.2±2.8 mmol·sec/L and 30.1±4.1 mmol·sec/L for stress and rest. No significant differences were found against automated results (AIF peak Gd, P=0.71 for stress and 0.94 for rest; first-pass duration, P=0.52 for stress and 0.84 for rest; AUC, P=0.67 for stress and 0.86 for rest). The manually generated AIF [Gd] curves were used to compute perfusion MBF maps. No significant differences were found between flow values computed with manual and automated masks (P=0.52 for stress and 0.65 for rest).

*Perfusion flow mapping*

The adenosine stress was found to induce a significant increase in myocardial blood flow. Mean flow at stress and rest were 2.82±0.47ml/min/g and 0.68±0.16ml/min/g, respectively. *PS* at stress was 1.32±0.26ml/min/g and significantly decreased to 1.09±0.21ml/min/g at rest (P<1e-3). Blood volume was 12.0±0.8ml/100g at stress and 9.7±1.0ml/100g at rest (P<1e-9), indicating good adenosine vasodilation response. $V_{isf}$ was 20.8±2.5ml/100g (equivalent to 19.8±2.4% volume fraction) at stress and 20.3±2.9ml/100g (19.3±2.8%) at rest (BTEX $V_{isf}$, stress vs. rest: P=0.50). The equivalent extracellular volume fraction (ECV) in percentage with BTEX model (computed as $100\times(V_{isf}+V_p)/1.05$) was 26.2±2.3% for stress and 24.7±3.2% for rest in good agreement with previously reported values using T1-mapping methods [55]. The extraction fraction ($E$) computed from $PS$ and $F$ using the BTEX model was 0.87±0.08 at rest and dropped to 0.55±0.05 at stress. Figure 8 shows an example of pixel-wise maps of BTEX model for stress and rest acquisition. The Supporting information Figure F1 provides the histogram plot of all parameters of all cases within the ROI. The cohort mean of MBF standard deviation across myocardium was 0.45±0.04ml/min/g for stress and 0.13±0.04ml/min/g for rest, corresponding to coefficient of variation (CV) of 16% for stress and 19% for rest.

*Sensitivity to HCT*

Sensitivity of flow estimates to HCT was illustrated by performing the flow mapping with fixed hematocrit from 0.3 to 0.6 in steps of 0.05 and a nominal value of 0.42. Figure 9 plots the MBF estimated with fixed HCT vs. the nominal value. The mean percentage error (Figure 9) was from -3.8% at HCT=0.3 to 2.5% at HCT=0.6.

*Processing time*

The time for image reconstruction and flow mapping was measured for the typical protocol which acquired the AIF and 3 slices for 60 heartbeats. Typical processing times were 119±9s using an external networked Linux PC computer with 16 physical cores (Intel Xeon CPU E5-2640 v3 @ 2.50GHz, 128GB RAM). This time was measured from the end of MR data acquisition to the moment when all maps were received by the scanner host computer. The processing time was also measured to be ~3 mins using the scanner's image reconstruction computer. For the specific MR scanner used in this study, the image reconstruction computer had 16 cores with Ubuntu 12.04 (Intel Xeon E5-2658 @ 2.10GHz, 64GB RAM).

# Discussion

*In-line implementation of automated workflow*

The potential benefits of quantitative perfusion for the diagnosis of cardiovascular diseases have long been recognized, but its accessibility is still limited, partly due to the lack of a standardized and practical technical solution. The proposed inline perfusion mapping computed MBF maps and sent them back to scanner at the end of perfusion data acquisition. This eliminated the need to perform off-line analysis and simplified clinical workflow for perfusion quantification. Typical ranges of absolute MBF values may be established for different gender, age or pathological conditions. The ability to deploy identical software solution on multiple sites is also helpful to establish and conduct cohort studies for different diseases and patient groups. This will favor the clinical acceptance of perfusion quantification and establish it as a valid diagnostic tool in the toolkit of cardiac MRI. The ability to generate perfusion maps inline on the scanner may allow timely refinement of an ongoing study while patients are still in the scanner.

The present study was performed on a 3T scanner using the SIEMENS PRISMA model, but has also been used at 1.5T with the SIEMENS AERA [34] and AVANTO FIT, and older generation 3T models such as SIEMENS SKYRA [56–58] and SKYRA FIT. While the Gadgetron software is independent of vendor, the current sequence has been developed on the SIEMENS platform.

*Motion correction*

Motion correction is an essential step to achieve fully automated inline flow mapping. The majority of proposed techniques to align perfusion images are based on image registration, e.g. either rigid body [59,60] or non-rigid [61–64]. It is recognized the rapid contrast change during the bolus passage makes robust image registration difficult [61]. To overcome this issue, different strategies were proposed, including progressively applying registration on consecutive perfusion images [48], detecting image features and tissue boundaries for registration [65] and estimating model image to minimize contrast change [61,62,64]. Among the latest category to which the proposed algorithm belongs, independent component analysis was used in [64] to separate LV, RV and myocardium and derive a motion-free reference. Principle component analysis was used in [62], where the registration was formulated as a two-step process. First the rigid body registration was used to remove the bulk motion and a non-rigid refinement step was added to align myocardium. This method required manually cropping a ROI around the heart to start the bulk motion correction step. Another model-based method utilized the two compartment kinetic model [66] to estimate perfusion response and motion correction was performed between estimate response image series and acquired images. There is no consensus at this moment as to the best approach. The proposed method, as compared to [62], has the advantage of being fully automated and did not require a rigid body registration step, because iterative model computation gradually removes large respiratory motion. Its disadvantage is the elevated computational cost. As a limitation of this study, proposed MOCO algorithm was not compared to other published methods. Instead, the focus of this paper was on the inline processing workflow and its overall performance.

*Comparison of CMR perfusion quantification with PET*

An independent comparison of our inline solution with BTEX model has been reported in [39]. This study compared perfusion flow mapping with 13N–NH3 cardiac positron emission tomography (PET) as a clinically accepted technique. Twenty-one patients were enrolled and underwent both CMR and PET perfusion scans on the same day. Excellent agreement between our CMR myocardial perfusion mapping method and PET perfusion was found for MBF (r=0.92 for global and r=0.83 for regional MBF) as well as global and regional myocardial perfusion reserve (MPR).

Repeatability of the proposed CMR myocardial perfusion mapping was tested and reported in [67]. Forty-two volunteers underwent repeated adenosine stress and rest perfusion CMR on the same day and again after a minimum interval of 7 days. There were no significant differences in intra- and inter-study stress and rest MBF or MPR. Within subject coefficient of variation was 8% for rest and 11% for stress within the same day, and 11% for rest and 12% for stress for studies repeated after 7 days. These values are comparable to those in the PET literature.

*Models and extraction fraction*

The principle to estimate myocardial blood flow (MBF) from [Gd] concentration utilizes the dynamics of Gd transport across the capillary membrane from the vascular space to the interstitial space. An in-depth review of this topic can be found at [11]. A number of models with varying complexity have been proposed to estimate MBF [13,25,50,68–70], and have been compared in several studies [9,19]. Models in these studies have included single compartment such as exponential [5], Fermi [7], model-free [8], 2-compartment exponential [6] and 2-compartment Fermi [71], and distributed parameters [9,11,13]. The basic concept of MBF estimation is to find the best fit model parameters such that when the forward model is applied to the AIF, the resulting myocardial signal will agree best with the signal in the least squares sense. An important distinction between models is their treatment of blood flow from the capillaries to the interstitium, also known as extravasation. Simpler models that estimate flow by deconvolution measure the unidirectional influx perfusion constant rate of Gd from blood space to the interstitial space, known as $K_i$ or also as $K_{trans}$. Rest perfusion has lower myocardial flow with higher extraction of Gd to interstitium. In this flow limited regime, $K_i$ is a good estimate of flow. Under adenosine stress condition, the extraction fraction is significantly reduced, and $K_i$ underestimates the MBF. Distributed models explicitly estimate additional parameters including the permeability surface (PS) area product that allow calculation of the extraction fraction. Distributed models also may estimate the interstitial volume and myocardial blood volume (MBV) which may potentially have additional diagnostic value.

There is not a current consensus in the literature on perfusion models or the best strategy for quantifying MBF. Our previous paper [34] presented comparison of BTEX to Fermi model, showing the latter gave slightly lower estimates of MBF. Since this paper is to present an algorithmic method of inline processing, detailed model comparison is out of the scope here. On the other hand, although only the BTEX method was implemented and tested in this study, the proposed inline mapping solution can act as an easy-to-use testing platform for different models, because prior processing steps, such as motion correction, Gd conversion, and surface coil inhomogeneity correction are already built into the workflow. It is possible to use the exported Gd images to test new flow models as well. This will allow a more rigorous comparison among different models by feeding them identical input data.

Results show stress flow had increased variation compared to the rest. This finding is consistent with myocardial flow quantification using PET [72] where 23 published PET studies were summarized for a total of 363 healthy volunteers. Some compounding factors may contribute to this increased variation. The imaging was more challenging for stress condition with elevated heart rate. Stronger breathing can lead to more through-plane motion and fluctuation of perfusion signal intensity. The sensitivity of perfusion flow modelling may decrease as the flow goes higher. This may lead to reduced accuracy of parameter estimation. Further analysis of these contributing factors remain an important research topic for perfusion flow quantification.

*Dependence on HCT*

The implemented BTEX model and other distributed models [11] require knowledge of the hematocrit. The models are nonlinear in nature and therefore the myocardial blood flow estimate does not scale linearly with input concentration, [Gd], which depends on the value of HCT. In this study, we varied the assumed HCT and compared results to the MBF estimated with measured HCT. For the normal subjects, the range of actual HCT was small and led to insignificant changes (within 1%) in estimate of blood flow when using an assumed nominal value of 0.42. Although it is possible to modify the workflow to enter the measured HCT, this is not always available at the time of scanning. The estimate of interstitial blood volume (similar to ECV) depends linearly on HCT, therefore, an estimate of the actual HCT is required to output this variable. An alternative strategy to directly measuring the HCT is to measure the pre-contrast T1 values of LV blood pool and estimate the HCT using the linear relationship of longitudinal blood relaxivity to the blood HCT, which has been demonstrated for synthetic ECV mapping [73]. Implementing this in-line without entering measurements from previous imaging would require a sequence modification to integrate a T1 measurement at the start of scan with longer trigger delay (TD) suitable for native blood T1. The influence of HCT was shown to be limited for the MBF parameter. The reason lies in equation 5, where the AIF Gd concentration is scaled up to be plasma concentration. The scaled AIF signal is input into the model to estimate plasma flow. The resulting plasma flow is scaled again for myocardial flow. The effect of two scaling steps tend to cancel, but not perfectly due to the non-linear nature of model. The same effect happens for blood volume which is less influenced by HCT. Other parameters, PS and interstitial volume, are more vulnerable to change in HCT, because of AIF input is scaled directly.

This study used the nominal HCT for the BTEX modelling, which is equivalent to assuming that the myocardium capillary HCT is the same as AIF blood. This simplifying assumption was used in previous publications [6,9,11,74] and also adopted here. The hematocrit of the capillary blood can be 63%–75% of HCT in large vessels [75,76] and difficult to measure in-vivo. This study showed MBF is not very sensitive to HCT, with the error being less than 5% for a wide range of supplied HCT values.

*Imaging technique: variation and limitation*

The dual-sequence implementation used in this study was designed to support both B-SSFP and Flash readouts. This study utilized the Flash readout at 3T, while B-SSFP is commonly used at 1.5T to improve the SNR. The inline processing was implemented for both sequence types. The current imaging protocol acquires 3 slices which does not provide good coverage of the apex. Greater coverage would be highly desirable. One approach to achieving this is to image every other heart beat allowing 6 slices to be imaged [56]. Importantly, the dual-sequence samples the AIF every heartbeat. This approach required reliable ECG gating.

This study utilized multi-slice 2D imaging with single-shot readout. The method is robust against arrhythmia and benefits from excellent saturation recovery [34]. One limitation is the lack of ability to capture through-plane motion. While non-rigid motion correction can correct respiratory motion, significant through-plane motion cannot be easily corrected in current scheme. Therefore, good slice planning is needed. To this end, 3D perfusion imaging may be desirable [77] or use of a navigator with slice tracking for prospective motion correction.

There are remaining technical problems to be solved for quantifying 3D perfusion imaging, including fast and robust 3D motion correction, intensity to Gd concentration conversion and pixel-wise flow mapping for large 3D volume.

*Validation*

Evaluation results were presented in this paper to verify the effectiveness of proposed technical algorithms. Quantitative results were presented for motion correction, AIF detection and pixel-wise flow mapping, with validation on imaging sequence and intensity to [Gd] conversion presented in [34]. Based on the proposed inline solution, there are comparison study with PET cardiac perfusion [39], showing good agreement between MR flow measurement and commercial PET MBF software. While all these results positively support the proposed inline solution, the need for clinical validation of perfusion flow mapping remains for different disease conditions, clinical settings, and multi-center trials. This paper is intended to serve as a technical starting point of introduction of inline flow mapping. The proposed fully automated solution may facilitate the clinical validation of perfusion flow mapping on a larger data cohort.

## Conclusions

We propose an automated workflow for inline quantitative perfusion flow mapping. The proposed solution allows free-breathing perfusion imaging and automated generation of myocardial flow maps without any user interaction. A distributed parameter Gd kinetics model (BTEX) was implemented in the proposed solution and tested on normal volunteers. This initial evaluation demonstrates the fully automated nature of the proposed solution and serves as the basis for further clinical validation.

# Appendices

*Motion correction*

MOCO used non-rigid image registration [45,46] applied in an iterative fashion. Source images were registered pairwise with target images derived from a Karhunen–Loève (KL) transform which computed principal eigen-images with similar contrast. The KL transform was computed over a sliding temporal window. At each stage of the iteration, the target images were recomputed from the registered images with a decreased window width. This approach can deal with significant contrast changes during the bolus passage which has been shown to be a major challenge for myocardial perfusion motion correction [47]. As shown in Figure 2, this algorithm iterates non-rigid image registration and KL transform based model image estimation to decouple perfusion contrast changes from respiratory motion.

*KL model image estimation*

Given a temporal window width $2W + 1$, a KL model image series can be derived from the perfusion image series. Assume a series of perfusion images as $f(i,t)$, where $i = 0,1, \dots, N_x N_y$ and $t = 0,1, \dots, N_t$. $N_x$ and $N_y$ are number of image pixels along readout and phase encoding direction. $N_t$ is the number of perfusion frames acquired in the scan. For a perfusion frame $t \in [0, N_t]$, a data matrix $\boldsymbol{f}_t$ can be assembled to include all frames from $[t - W, t + W]$:

$$\boldsymbol{f}_t = \begin{bmatrix} f(0,t-W) & f(0,t-W+1) & \dots & f(0,t+W) \\ f(1,t-W) & f(1,t-W+1) & \dots & f(1,t+W) \\ \dots & \dots & \dots & \dots \\ \dots & \dots & \dots & \dots \\ f(N_x N_y,t-W) & f(N_x N_y,t-W+1) & \dots & f(N_x N_y,t+W) \end{bmatrix} \quad (A.1)$$

A KL eigenimage can be computed by multiplying the leading eigenvector corresponding to the maximal eigenvalue of $\boldsymbol{f}_t$. This process was repeated for all $N_t$ frames in a sliding fashion, to create a KL model image series $M$. Unlike the simple averaging across the sliding window $W$ or other low pass filter, the KL eigenimage is the optimal low-rank approximation of data matrix using the minimal least-squares criteria. It keeps the most prominent image information corresponding to the first eigenmode and filters out the respiratory motion which is assumed to be continuous and sampled sufficiently within the temporal window. Since the perfusion images are usually acquired every one or two heart beats, the respiratory motion is sampled with sufficient temporal resolution to fulfill this requirement. As demonstrated in Supporting information video 1, the model images with wider temporal window kept less temporal information, but filters out respiratory motion. The output of the first iteration removed bulk respiratory motion, but some residual motion remained. As the model window is narrower, more temporal contrast changes were preserved in the model series. An updated series of model images at each iteration was computed from the previous MOCO output, so more residual motion can be corrected after each iteration.

*Non-rigid image registration and iterative motion correction*

The original perfusion series was registered to the KL model image series in a frame-by-frame manner. Because of the non-rigid nature of motion presented in the field-of-view (FOV), a non-rigid registration algorithm was applied [45,46] to maximizing local cross-correlation as the registration cost function. The outcome of this algorithm is the pixel-wise deformation field indicating the motion vector of every image pixel. This algorithm is based on variational theory and modelled the deformation field as an unknown functional to maximize the image similarity measure between KL model and perfusion images. The classic gradient descent method was used to solve the corresponding Euler equation for the optimal functional. To maximize the capture range and improve the robustness, a multi-scale image pyramid was constructed by downsampling the images (4 levels in all experiments with 2x downsampling at each level). The deformation fields estimated on the coarse scale were used to initialize the finer level, until the original image resolution was achieved. A maximum of 64 iterations were performed on every scale level until the image similarity measure reaches its maximum. The local cross correlation ratio [45] is selected as the image similarity measure, as its explicit derivative can be effectively calculated, which is used in gradient descent optimization, and is still general enough to handle image noise and the remaining intensity changes between the KLT model images and the perfusion images series.

As shown in Figure 3, during the first iteration, the KL model images ($M_0$) were estimated with a wide window $W_0$ and registration was performed between this model series $M_0$ and perfusion images $f$. $f$ was warped with the resulting deformation fields, leading to a new image series with less respiratory motion. This process was repeated by computing the new KL model image series $M_1$ on the warped perfusion images with a narrower sliding window $W_1 = W_0/2$ until reaching the narrowest window $W_{min}$. $W_0 = N_t/3$ and $W_{min} = 3$ were experimentally chosen and found to give very robust motion correction results. For a perfusion acquisition lasting 60 heart beats, this led to a total of 4 iterations for KL model series estimation and non-rigid image registration. This scheme of iterative MOCO setup provides an empirically good capture range for non-rigid registration and can adapt to longer acquisitions. Co-registration between model series and perfusion series was performed frame by frame since the model frame had approximately the same image contrast as the target perfusion images. Since at each iteration all image pairs were processed independently from each other this avoided error propagation and permitted utilizing multi-threading in the Gadgetron framework to speed up computation.

This motion correction algorithm was applied to both the AIF and higher resolution myocardial perfusion images. In the case of the AIF, MOCO was applied to the first echo AIF image series and the resulting deformation fields were used to correct the second echo images. After the perfusion image series were aligned, the last 6 images were averaged as a reference to further align the PD images.

*Perfusion mapping*

Pixel-wise perfusion flow mapping is calculated by fitting the distributed blood tissue exchange (BTEX) model [36,50] to the measured myocardial perfusion [Gd] signal independently for each pixel in the heart region. In this study, a numerical partial differential equation (PDE) solver [78] was implemented to compute the $C_p(t,x)$ and $C_{isf}(t,x)$. The evaluated capillary length ranges from 0 to L (set to be 1mm as in [37]) and the evaluated temporal duration is the full duration from arrival of Gd bolus to LV (detected foot time) through the end of imaging acquisition. The initial conditions are $C_p(0,0) = 0$ and $C_{isf}(0,0) = 0$ and the boundary condition is set as the AIF input $C_p(t,0) = C_a(t)$. The residual Gd over time $Q(t) = F_p \cdot \int_0^t [C_a(s) - C_p(t,L)]ds$ is the accumulated contrast agent in the system. As suggested in [37,51], we choose to vary $F_p$ (ml/min/g), $PS$ (ml/min/g), $V_p$ (ml/g) and $V_{isf}$ (ml/g).

Figure 5 demonstrates the iterative process for BTEX modelling. The fitting process starts with an initial guess of model parameters. The corresponding partial differential equations in BTEX model are solved with the AIF [Gd] curve as the driving input function. The resulting Gd residual signal is compared to the measured perfusion [Gd] curve for the computation of the mean squared error (L2 norm) as the goodness-of-fit measure. The BTEX parameters are then adjusted in the optimization step. This process iterates until the convergence. The step that is computationally expensive is applying the partial differential equations (PDE) to the AIF which is the forward model. Parameter estimation is done in coarse and fine steps. The coarse step does a brute force search over the full parameter range with relatively coarse steps ($F_p$, from 0.1 to 3ml/min/g with 0.05 step size; $PS$, from 0.4 to 1.8ml/min/g with 0.1 step size; $V_{isf}$, from 0.15 to 0.65ml/g with 0.025 step size; $V_p$, from 0.035 to 0.08ml/g with 0.005 step size; Note the plasma flow and volume is used here, i.e., without HCT). The parameter search ranges were selected to cover the expected full span for the myocardium [9,79,80]. In this step, the PDE is applied to the AIF for all searching parameters sets (185,850 sets in total). The computed myocardial [Gd] response signal was stored and compared to the measured signal for every pixel. The coarse search setup finds its answer by picking the parameters corresponding to closest [Gd] response signal to the measured one in the minimum least-square-error sense. This is followed by a fine search step with an iterative optimization initialized by the coarse search setup. Since computing analytical derivatives of BTEX model to its parameters is nontrivial and evaluating numerical derivatives is also computationally expensive, the downhill simplex minimization algorithm proposed by Nelder and Mead [81] is used for the final optimization step, as it is more robust for non-smooth cost function and does not require evaluation of parameter derivatives. The coarse/fine parameter estimation strategy is diagrammed in Figure 10.

Since the AIF signal is measured in the LV blood pool (at the most basal slice of perfusion imaging stack), there is an unknown time delay for the contrast agent to reach the myocardium. Different approaches have been proposed to compensate for this effect, including fitting time delay as an extra parameter [71] or assuming a constant wash-in time [9,19]. In this study, we adopted a multi-fitting approach that fits at multiple values of delay in 0.5 s increments and chose the delay with best fit. A maximal delay of 3.0s was allowed. The estimate of delay was done independently for each pixel. The incremental delay range was picked empirically to balance computational cost and fitting accuracy.

**Abbreviations**

| | |
|---|---|
| AIF | arterial input function |
| AUC | area-under-curve |
| BTEX | blood tissue exchange model |
| Dice | Dice similarity coefficient |
| DP | distributed parameter |
| E | extraction fraction |
| ECV | extracellular volume fraction |
| FA | flip angle |
| Fermi | constrained Fermi function |
| FLASH | fast low angle shot |
| FOV | field-of-view |
| HCT | hematocrit |
| [Gd] | gadolinium concentration |
| KL | Karhunen–Loève |
| LGE | late gadolinium enhancement |
| LUT | look-up table |
| MBF | myocardial blood flow |
| MPR | myocardial perfusion reserve |
| MOCO | motion correction |
| PD | proton density |
| SNR | signal-to-noise ratio |
| SR | saturation recovery |
| SSFP | steady state free precession |
| TD | saturation recovery delay time |
| TS | saturation time |

**Declarations**

**Ethical Approval and Consent to participate**

Imaging was performed at the University of Leeds Hospital, Leeds, UK. Studies were approved by the local Ethics Committees and written informed consent for research was obtained for all subjects. Anonymized data were analyzed at NIH with approval by the NIH Office of Human Subjects Research OHSR (Exemption #13156).

**Consent for publication**

Written informed consent was obtained from patients for publication of their individual details and accompanying images in this manuscript. The consent form is held in the patients' clinical notes and is available for review by the Editor-in-Chief.

**Availability of data and material**

The raw data that support the findings of this study are available from the corresponding author upon reasonable request subject to restriction on use by the Office of Human Subjects Research. Raw data require reconstruction processing. The BTEX flow mapping software is shared at a github repo (*https://github.com/xueh2/QPerf.git*). An example perfusion dataset can be downloaded from this repo and for testing myocardial blood flow estimation using the BTEX model provided. A detailed user's guide is provided.

**Competing interests**

The authors declare that they have no competing interests

**Funding**

Supported by the National Heart, Lung and Blood Institute, National Institutes of Health by the Division of Intramural Research.

**Authors' contributions**

HX and PK conceived of the study and developed the algorithms, implemented the inline reconstruction and processing software, performed processing and analysis, and drafted the manuscript. PK, HX, and SNV developed the sequence. LB and SP acquired normal volunteer data. All authors participated in revising the manuscript and read and approved the final manuscript.

**Acknowledgements**

We acknowledge Prof. James B. Bassingthwaighte (Univ. of Washington) and his team for his inspirational work on the blood tissue exchange model for myocardial perfusion.

**Additional files**

**Supporting Information Video S1:** An example of KL model-based motion correction, showing the model series and motion corrected series for every KL iteration (corresponding to Figure 3).

**Supporting Information Video S2:** Movie of a perfusion scan demonstrates the free-breathing acquisition with heart moving with respiratory (left: stress scan; right: rest scan), corresponding to Figure 7a-b.

**Supporting Information Video S3:** After motion correction, the spatial alignment of myocardium is restored, and pixel-wise mapping becomes feasible (left: stress scan; right: rest scan), corresponding to Figure 7e-f.

**Supporting Information Video S4:** Dual-echo AIF image series after motion correction. The detected LV mask is overlaid on the first echo images.

**Supporting Information Video S5:** An example of motion correction for a score of 4.0 (good). First row is the original images with free-breathing acquisition and the second row is motion corrected images. Both stress (column 1-3) and rest (column 4-6) are given.

**Supporting Information Video S6:** An example of motion correction for a lower score of 3.0.

**Supporting Information Figure S1**: The histogram of fitting parameters for the tested data cohort.

# List of Captions

**Figure 1** Overview of the proposed workflow for automated inline perfusion flow mapping. After raw kspace data are reconstructed, resulting low resolution AIF and high resolution perfusion images go through the motion correction step, which allows the free-breathing acquisition. The AIF image series was inputted to LV blood pool detection and resulting signals were corrected for T2* signal loss. The surface coil inhomogeneity was corrected by normalized perfusion series with proton density images. Both T2* corrected AIF signal and normalized MOCO perfusion images are converted into [Gd] unit by a look-up-table conversion. Finally, AIF Gd curve and perfusion Gd images are inputted into flow mapping step for pixel-wise myocardial flow mapping.

**Figure 2** A flow chart of perfusion motion correction scheme with iterative KL transform based model image estimation. Starting with a wider temporal window, this algorithm iterates KL model estimation and pair-wise image registration between model and original series. This decouples image contrast changes from respiratory motion correction. This MOCO scheme is applied to both AIF and perfusion images and generate motion corrected image series.

**Figure 3** An example of KL based motion correction. The original free-breathing perfusion series are shown on the top row. The first and third iteration of KL model based MOCO is also shown for model series (M0 and M2) and MOCO outputs (f0 and f2). Respiratory motion is recovered with narrower temporal window after MOCO iterations (Corresponding movie in S1).

**Figure 4** An illustration of AIF LV blood pool detection. The LV blood pool is detected by first thresholding the upslope and AUC maps to find a rough mask of heart. The LV blood pool is delineated by a two-stage clustering process to compute arterial input signal.

**Figure 5** An illustration of iterative pixel-wise perfusion flow mapping. Inputs to this fitting process are the AIF Gd signal and perfusion Gd images. By iteratively solving the BTEX equations, the model

parameters are adjusted to reduce the discrepancy between the estimated and the measured Gd signals. This fitting process is performed for every pixel and resulting perfusion flow maps are computed.

**Figure 6** Proposed perfusion flow mapping was integrated on the scanner. (a) is a screenshot of inline perfusion flow mapping scan for a patient with obstructive epicardial coronary artery disease for illustration of the method. The pixel-wise MBF map, AIF figures and perfusion motion correction images are sent back to scanner without any user interaction. (b) gives example AIF plots for stress and rest scans. The AIF intensity curves of dual-echo acquisition are shown as the first column. The middle column is the AIF curve in Gd unit. The last column plots RR intervals which was clearly reduced in stress, compared to the rest scan.

**Figure 7** An example to demonstrate typical performance of motion correction. Original perfusion series of stress (a) and rest scan (b) is acquired under free-breathing. (c-d) Temporal profiles before MOCO show respiratory motion across different images (Supporting information video S2). After the motion correction (e-f), the heart is aligned during the contrast uptake, which allows the pixel-wise flow mapping (S 3). The temporal profile after MOCO (g-h) shows the removal of heart motion due to respiratory.

**Figure 8** An example of perfusion flow mapping. Pixel-wise maps for all four parameter and extraction fraction are computed by the proposed automated workflow. The stress flow and blood volume are significantly increased compared to rest. The extraction is higher at rest and lower at stress, indicating the stress myocardium is not flow-limited. Given the ROI drawn in the myocardium, histograms of all parameters for entire cohort are given for stress and rest mapping in the Supporting information figure S1.

**Figure 9** Influence of hematocrit on the myocardial blood flow estimation. The MBF estimated with fixed HCT is compared to those estimated with the nominal hematocrit 0.42. For a range of hematocrit from 0.3 to 0.6, the mean variation of MBF is less than 4%.

**Figure 10** A flow chart for the coarse/fine parameter estimation strategy. The BTEX model fitting starts by a brute force search over the full parameter range with relatively coarse steps. An optimal starting point is found in the coarse step and used to initialize the fine search.

# Figure 1. Overview of workflow.

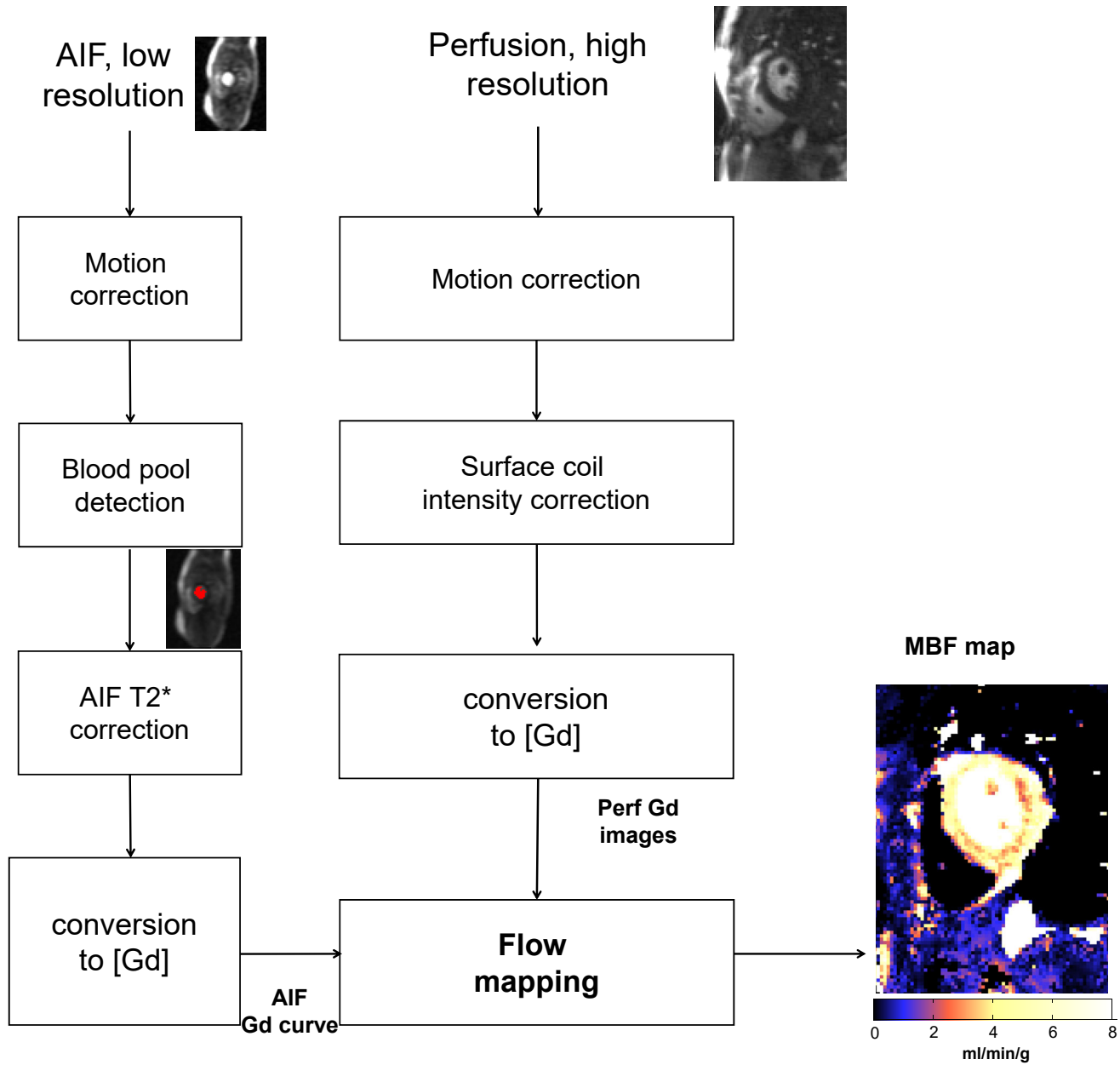

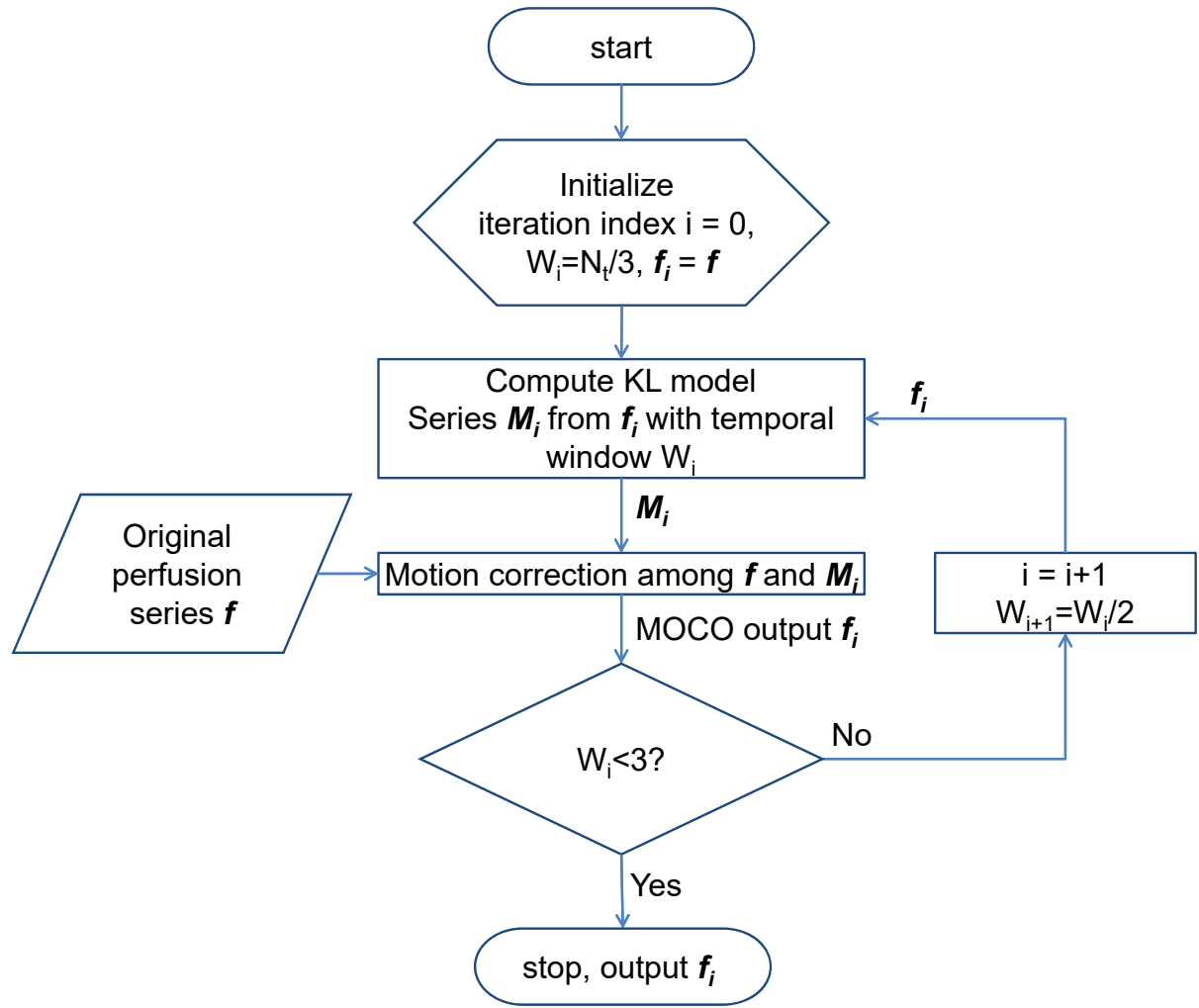

Figure 2. Flow chart of proposed KL model based perfusion motion correction algorithm.

**Figure 3.**

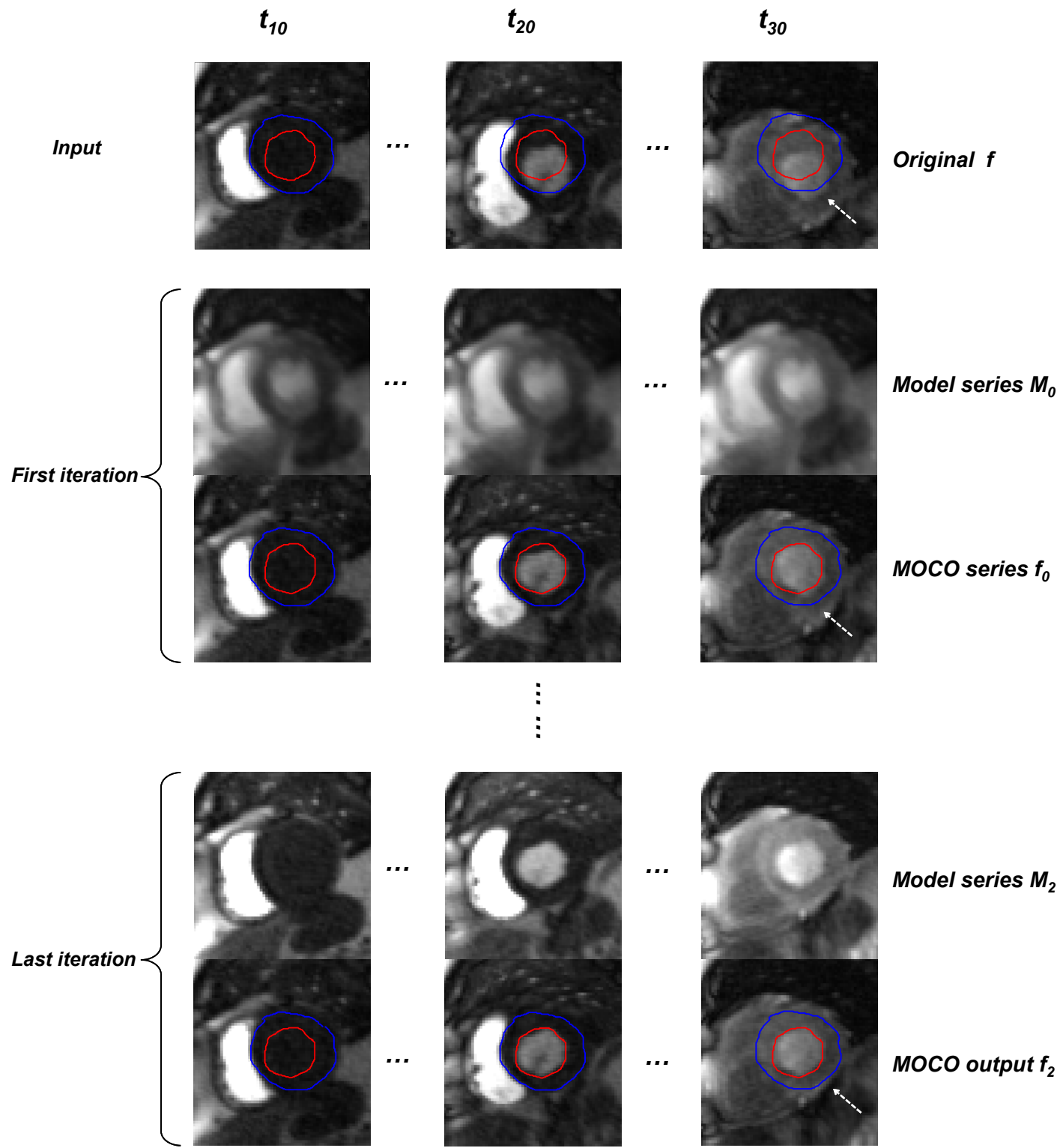

**Figure 4.**

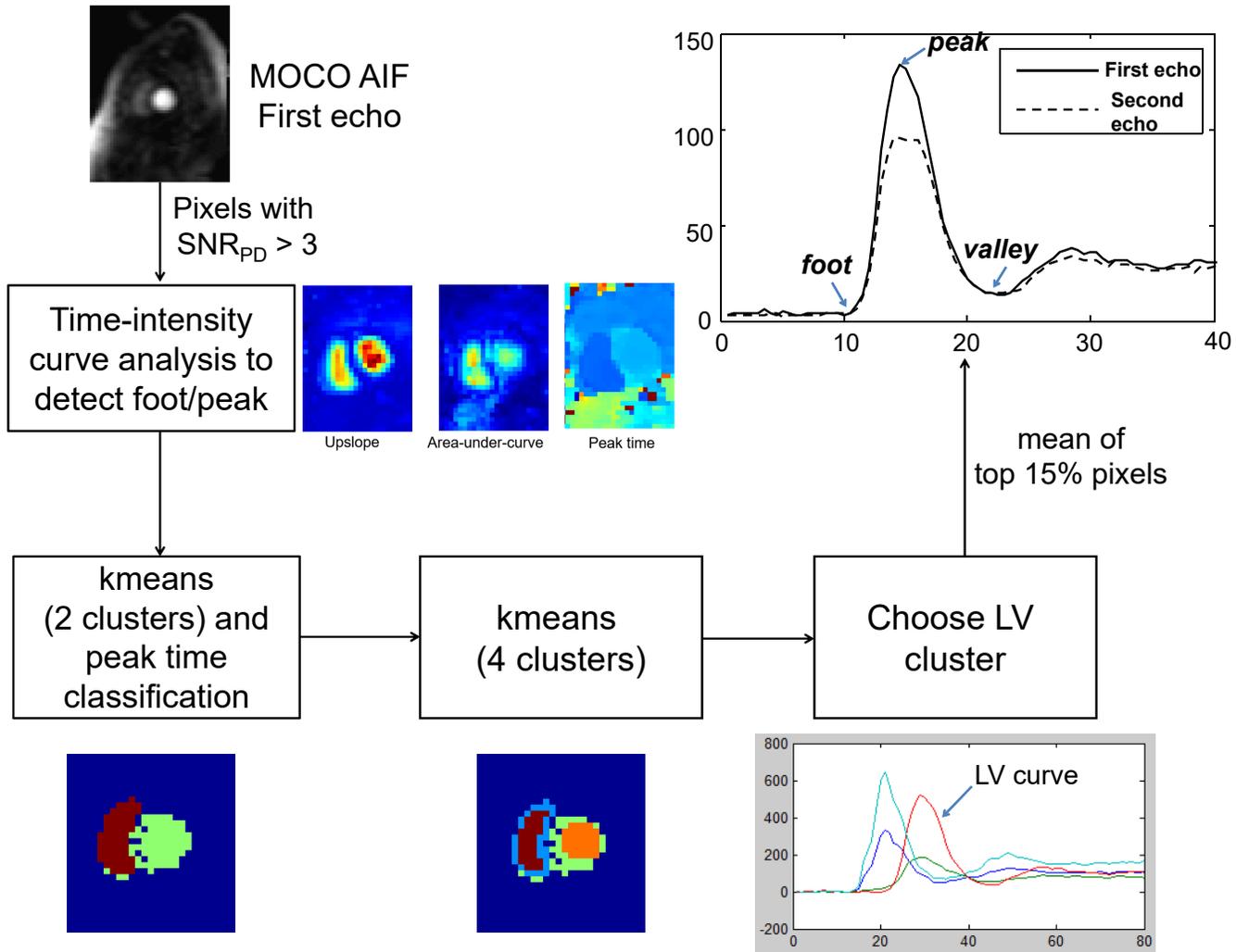

**Figure 5.**

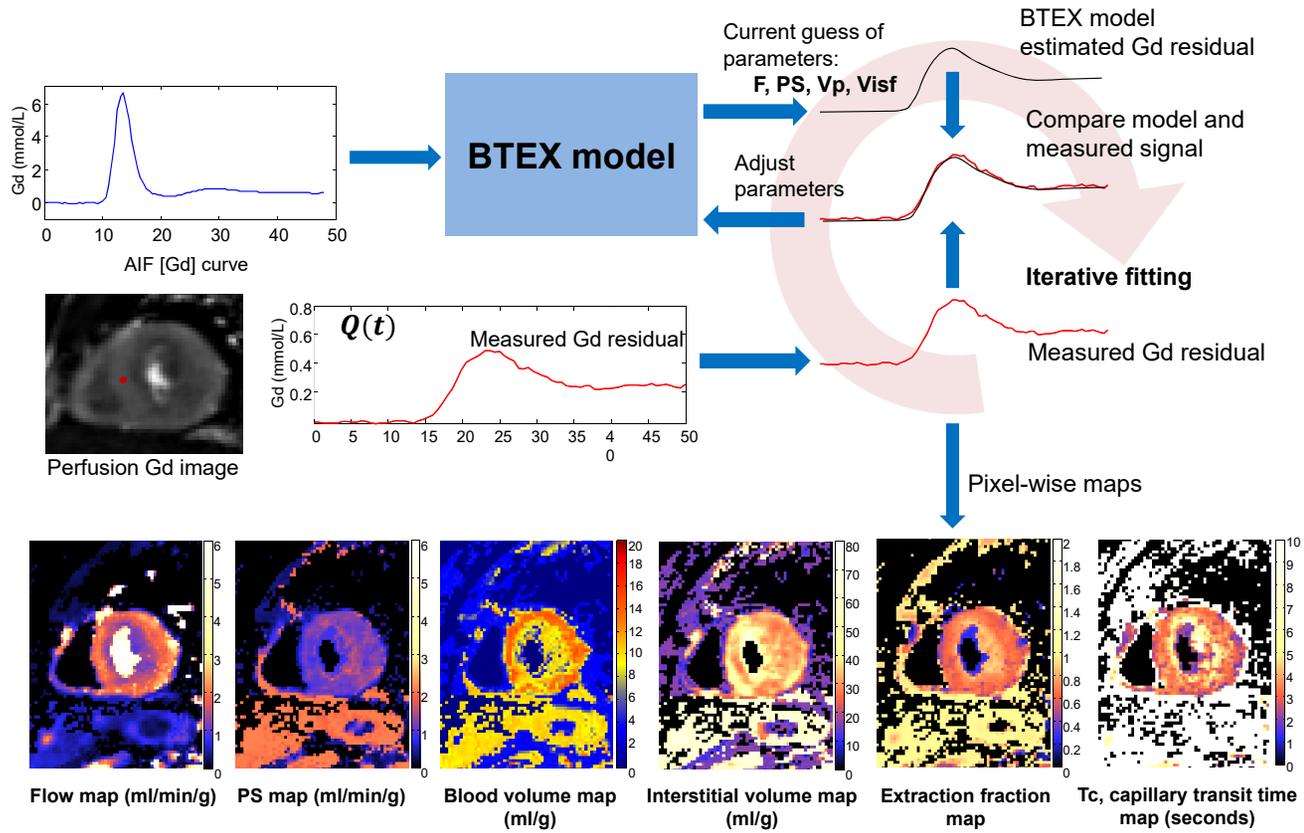

**Figure 6.**

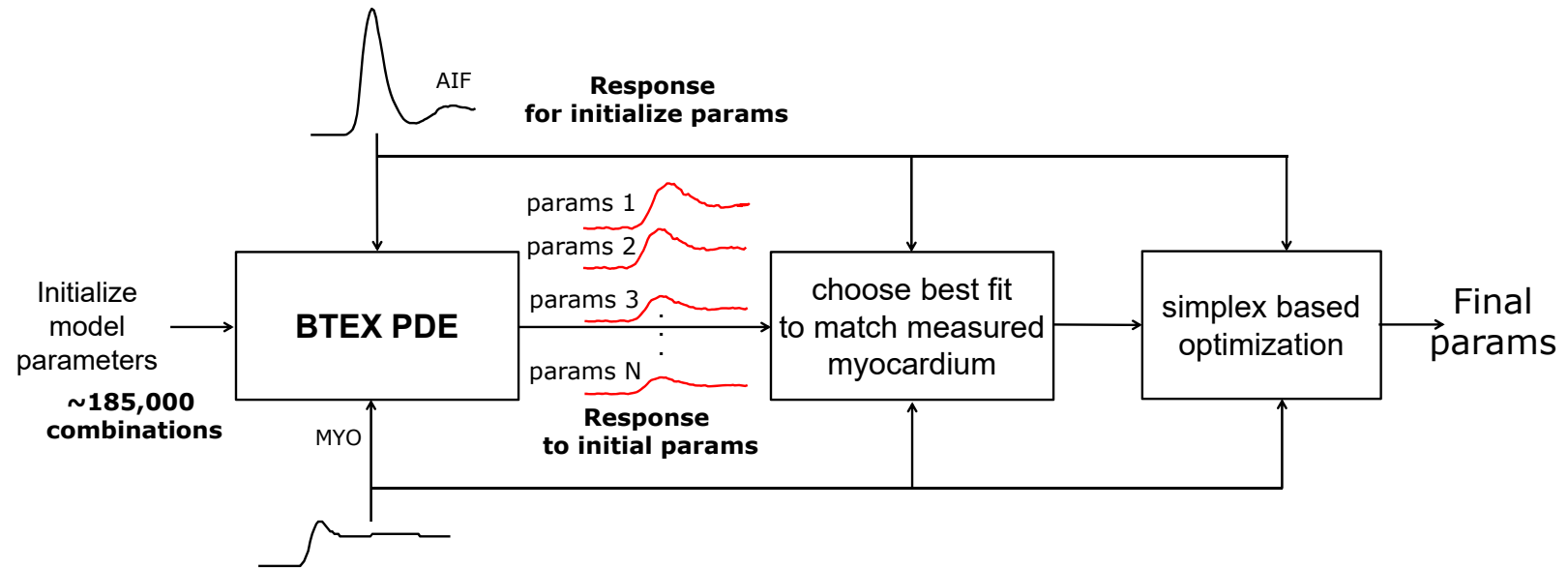

# Figure 7

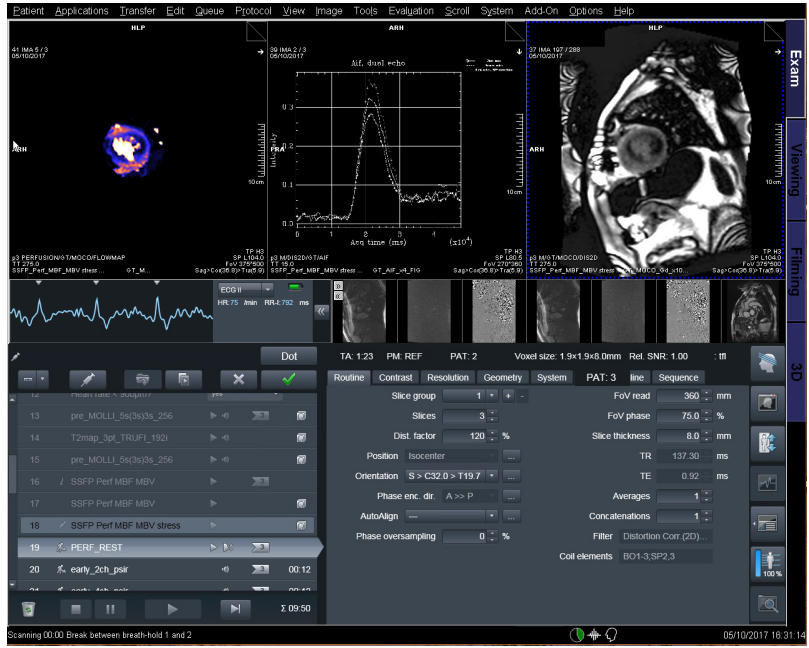

(a) Screenshot of inline perfusion flow mapping scan

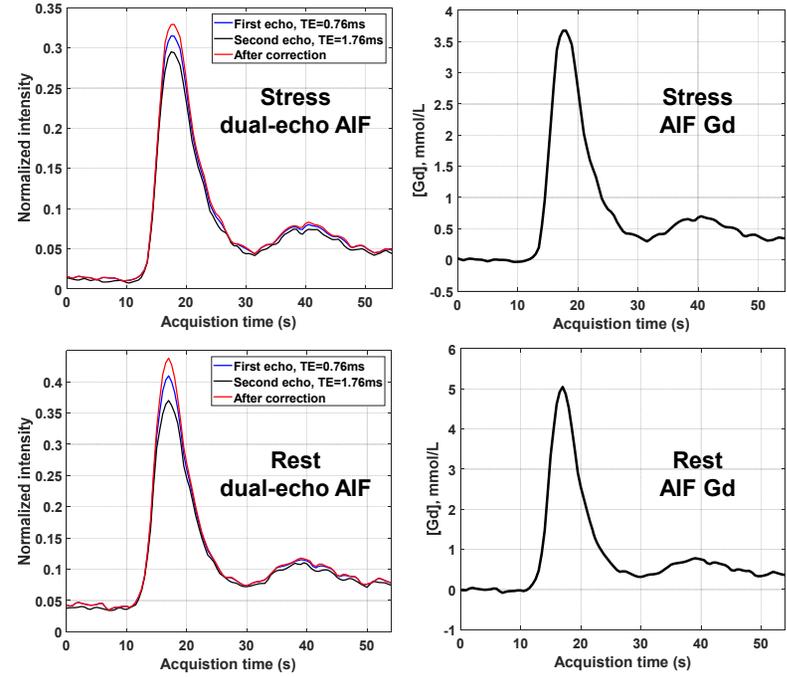

(b) Example AIF plots of stress and rest scans

**Figure 8**

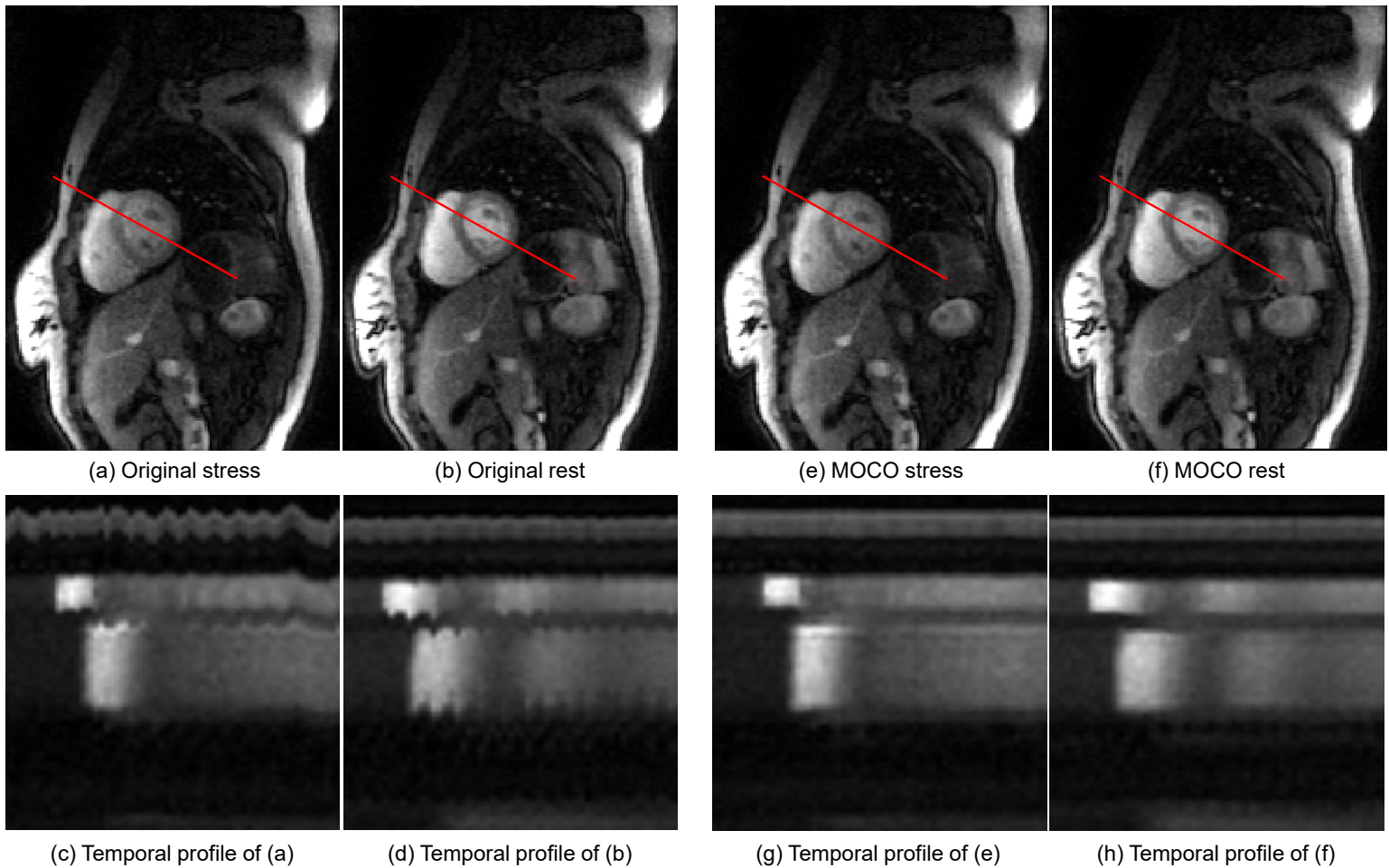

Figure 9. An example of perfusion flow mapping.

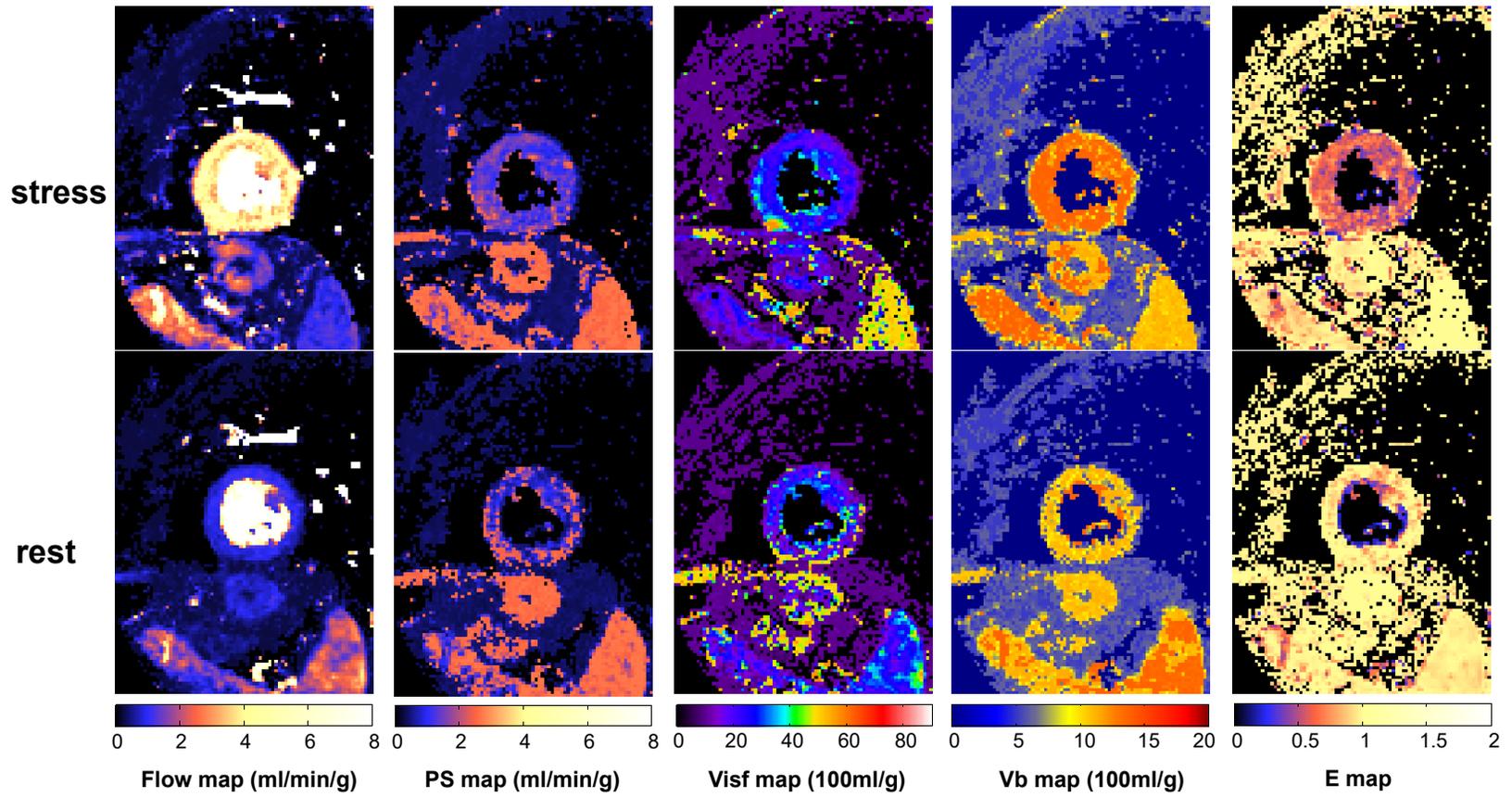

**Figure 10**

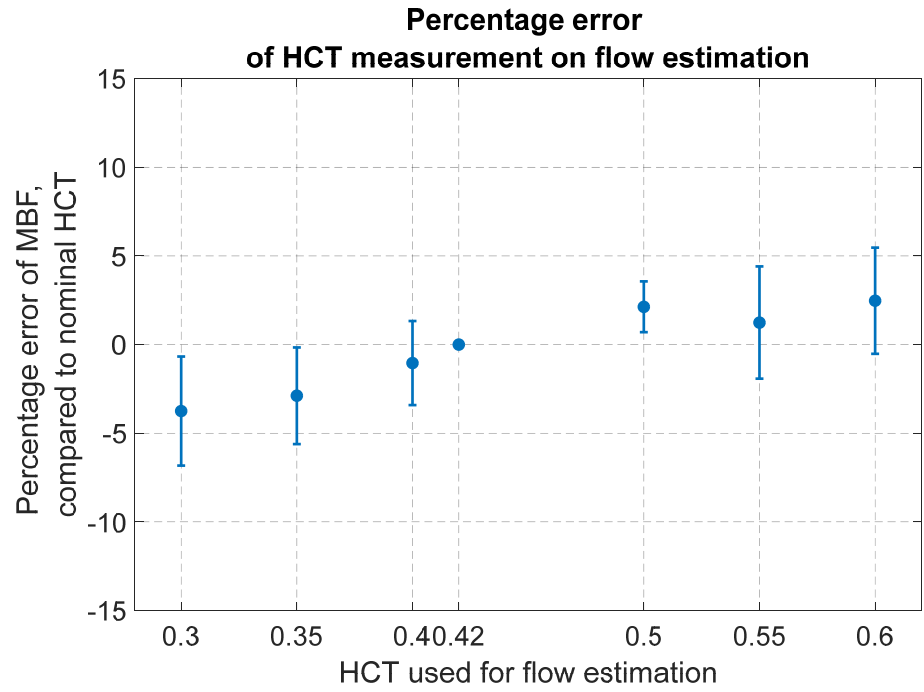

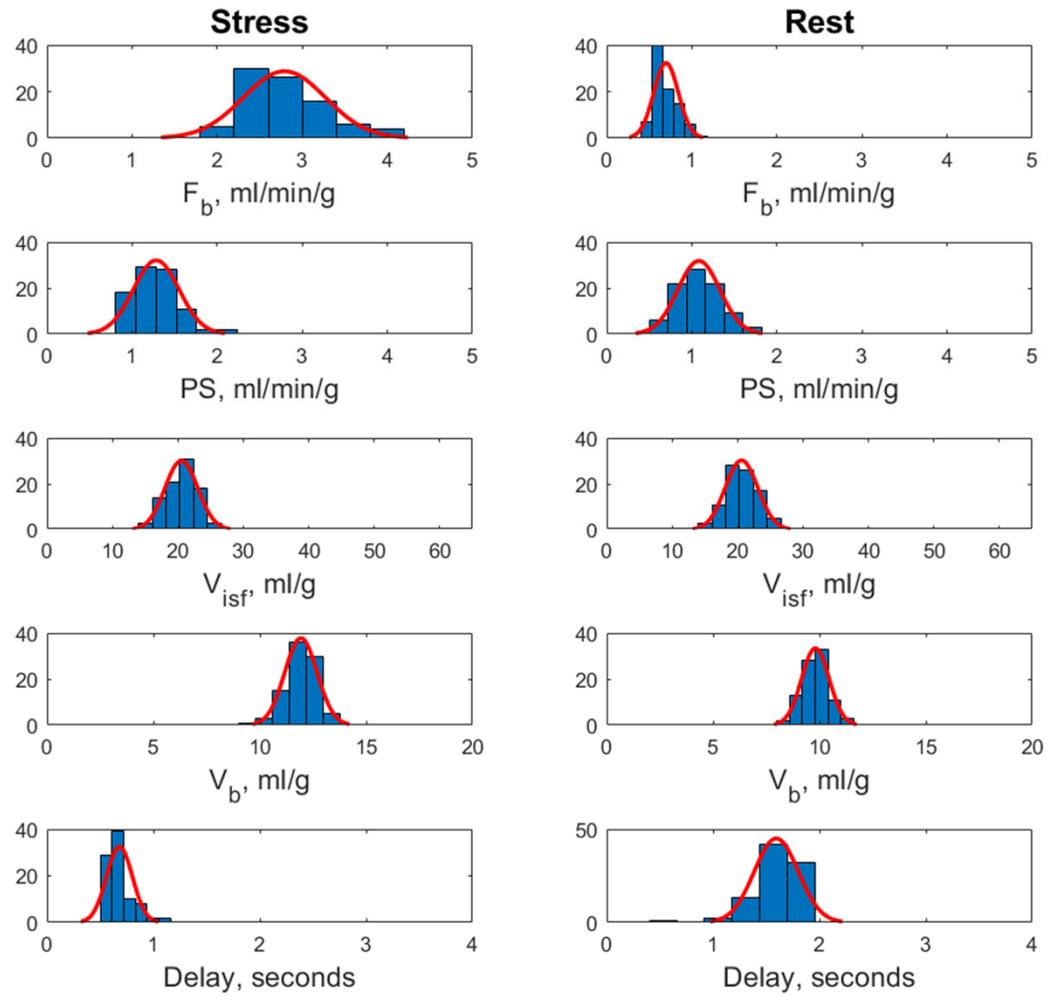